\documentclass[
superscriptaddress,
nofootinbib,
amsmath,amssymb,
longbibliography,
twocolumn %remove for preprint
]{revtex4-2}

\usepackage{wrapfig}
\usepackage{graphicx}
\usepackage{lineno}
\usepackage[utf8]{inputenc}
\usepackage{rsfso}
\usepackage{amsmath,color}
\usepackage{graphicx}% Include figure files
\usepackage{dcolumn}% Align table columns on decimal point
\usepackage{bm}
\usepackage{amsfonts}
\usepackage{amsthm}
\usepackage{breqn}%break equations over 2 lines
\usepackage{xcolor}
\usepackage{hyperref}

\newcommand\beq{\begin{equation}}
\newcommand\eeq{\end{equation}}
\newcommand\beqa{\begin{eqnarray}}
\newcommand\eeqa{\end{eqnarray}}

\renewcommand{\figurename}{Figure}
\makeatletter
\def\fnum@figure{\textbf{\figurename~\thefigure}}
\makeatother
\makeatletter
\renewcommand\NAT@open{(}
\renewcommand\NAT@close{)}
\makeatother

\newcommand{\Du}{\textnormal{Du}}
\newcommand{\Pe}{\textnormal{Pe}}

\newcommand{\p}{_{\rm inlet}}
\newcommand{\ch}{_{\rm ch}}

\newcommand\LB[1]{\textcolor{black}{{#1}}}

\newcommand\ML[1]{\textcolor{black} {{#1}}}

\newcommand\Rev[1]{\textcolor{black} {{#1}}}
\newcommand\LBB[1]{\textcolor{black}{{#1}}}
\newcommand\QY[1]{\textcolor{black}{{#1}}}
\newcommand\RevEditor[1]{\textcolor{black} {{#1}}}

\def\Pe{\rm{Pe}}
\def\Du{\rm{Du}}

\def\Pe{\textnormal{Pe}}
\def\Du{\textnormal{Du}}

\def\.{\cdot}
\def\1{^{-1}}
\def\2{^{-2}}
\def\3{^{-3}}

\begin{document}

\title{Architecting mechanosensitive nanofluidic transport in graphite nanoslits}

	\author{Mathieu Lizée}
    \email{lizee@fhi-berlin.mpg.de}
    \thanks{Current address: Fritz Haber Institute of the Max Planck Society, Berlin, Germany}
    \affiliation{\rm These authors contributed equally}
    \affiliation{\rm Laboratoire de Physique de l'Ecole Normale Supérieure$,$ ENS$,$ Université PSL$,$ CNRS$,$ 24 rue Lhomond$,$ 75005 Paris$,$ France}

    \author{Zhijia Zhang} 
    \affiliation{\rm These authors contributed equally}
    \affiliation{\rm Department of Physics and Astronomy$,$ University of Manchester$,$ Manchester$,$ M13 9PL UK}
     \affiliation{\rm National Graphene Institute$,$ University of Manchester$,$ Manchester$,$ M13 9PL UK}
    \author{Baptiste Coquinot}
    \affiliation{\rm Institute of Science and Technology Austria (ISTA)$,$  Am Campus 1$,$ 3400 Klosterneuburg$,$ Austria}
    \author{Qian Yang}
    \email{qian.yang@manchester.ac.uk}
    \affiliation{\rm Department of Physics and Astronomy$,$ University of Manchester$,$ Manchester$,$ M13 9PL UK}
    \affiliation{\rm National Graphene Institute$,$ University of Manchester$,$ Manchester$,$ M13 9PL UK}
    
    \author{Lydéric Bocquet}
    \email{lyderic.bocquet@ens.fr}
    \affiliation{\rm Laboratoire de Physique de l'Ecole Normale Supérieure$,$ ENS$,$ Université PSL$,$ CNRS$,$ 24 rue Lhomond$,$ 75005 Paris$,$ France}

\begin{abstract}
Mechanosensitive ion transport plays a central role in enabling living systems to perceive and adapt to their environment through the deformation of soft, embedded ion channels. In this work, we demonstrate that ion transport within a two-dimensional graphite nanoslit can be rationally engineered to achieve a bipolar, pressure-sensitive response \Rev{without any structural deformation.} The mechanosensitivity arises from the selective charging of one channel inlet, which acts as a reversible \Rev{source of mobile charge carriers. 
These excess-ions can then be advected in or out of the channel by the pressure-driven water flow,  thereby modulating the ionic conductance.
%The experimental pressure dependence} 
This mechanism} is captured through a comprehensive electrohydrodynamic model that analytically accounts  for coupled diffusion, convection, surface transport, diffusio-osmosis, and interfacial slippage, both inside and outside the nanoslit. 
%The theoretical framework successfully reproduces the experimental data, %and reveals that spatially patterned surface charges markedly amplify the system's pressure sensitivity. %\Rev{These findings establish a route toward the design of fully-ionic nanofluidic pressure sensors free of any reliance on soft deformable materials which are notoriously hard to engineer at the nanoscale.} 
\LBB{
%demonstrating that spatially patterned surface charge pattern can induce complex pressure-dependent conductance. This reveals how 
%the rich non-linear couplings emerging  at the nanoscale open the way to adaptive, bioinspired reponsive nanofluidic systems, here in the form of ionic pressure sensors.
%or  soft and responsive nanofluidic systems.
The theoretical framework quantitatively reproduces the experimental data, showing that a simple surface charge pattern can give rise to complex, pressure-dependent conductance. These findings reveal how rich nonlinear couplings at the nanoscale can be harnessed to design adaptive, bioinspired nanofluidic systems, exemplified here by ionic pressure sensors.
}
\end{abstract}
\maketitle

\paragraph*{Teaser: } Local surface charge engineering enables ionic pressure-sensing in graphitic nanofluidic channels.

\RevEditor{\section*{Introduction}}
Living organisms experience a great variety of mechanical forces to which they adapt. Consequently, Nature has developed exquisite pressure sensors in the form of embedded ion channels across cellular membrane that deform under \Rev{membrane tension or differential pressure drops} producing strong current signals for pressure changes as low as 10~mbar \cite{wu2017touch, cox2019biophysical}. In parallel to the biological porins, solid-state nanopores and nanochannels offer new model systems with a high degree of structural and chemical control. In some nanopore geometries, strain-induced mechanical transduction was shown to \QY{mimic} some pressure sensitivity \cite{davis2020pressure,macha2022stress}.

However, beyond direct mechanical mimicry, the rich and intertwinned transport of ions and fluids inside nanochannels offers a wealth of possibilities to modulate ion conduction under various external stimuli \Rev{allowing to design sensing functions similar to those found in Nature, albeit based on distinct working principles}. Several experiments have reported non-linear coupled transport effects, taking the form of pressure modulation of the conductance or voltage-modulation of the streaming current \cite{jubin2018dramatic,mouterde2019molecular,marcotte2020mechanically,jiang2022nonlinear}. Depending on the geometry, such coupled behaviors were attributed to the modulation of the charge polarization in conical nanopores \cite{jubin2018dramatic,paul2024nanoscale}, or to the advection of potential-induced charges by the flow in carbon nanotubes or across graphene nanopores \cite{marcotte2020mechanically, jiang2022nonlinear}. 
%which is strongly enhanced for confining materials with low surface friction \cite{marcotte2020mechanically,mouterde2019molecular}. 
\Rev{Although it is now clear that }mechano-sensitivity can emerge via the coupling between ionic and hydrodynamic flow in the absence of mechanical deformation, \Rev{the lack of a clear mechanism and suitable design hinders progress towards strain-free nanofluidic pressure sensing.} Furthermore, while in the above systems, the effects of pressure were systematically symmetric under pressure reversal ($\Delta P\rightarrow -\Delta P$), it would be desirable to design systems \QY{ that are} \Rev{able} to sense the sign of the pressure drop as well.

%effect of pressure on ion conduction has been 

%\LB{For example, the influence of pressure on the space charge region across conical nanopipettes was shown to lead to pressure dependent ion conduction\cite{jubin2018dramatic,paul2024nanoscale}, while for single-digit carbon nanotubes a quadratic dependence of the conduction on the applied pressure was evidenced and attributed to strong flushing associated with large slippage specific to the CNT water interface \cite{marcotte2020mechanically}. 
%\LB{ In graphene nanopores, a similar model based on the advection of potential-induced charges by the flow, was used to explain a symmetric and linear G(P) law \cite{jiang2022nonlinear,paul2024nanoscale}. \cite{mouterde2019molecular} }
%In solid-state channels, surface charge, asymmetric designs and water slippage at atomically smooth surfaces have led us to see new mechanisms unknown in soft biological systems. Such new phenomenology was observed in conical glass nanopipettes \cite{jubin2018dramatic}: the influence of pressure on the space charge region similar to that of PN junctions was shown to yield very high pressure sensitivity without mechanical deformation. The large slippage of the inner wall of small-diameter carbon nanotubes -- which can now be routinely embedded in lipidic membranes -- led to a parabolic dependence of conductance on pressure \cite{marcotte2020mechanically}. The origin of mechano-sensitivity is not structural deformation of the pore, but the coupling between ionic and hydrodynamic flows. 

In this work, we demonstrate that a strong mechanosensitive behavior of a graphitic nanochannel can be engineered through the deliberate design of surface charge patterns, giving rise to a nonlinear pressure-dependent ionic transport.
%In this paper, we show that one can engineer a mechanosensitive response of a graphitic nanochannel by making use of surface charge patterns and the resulting non-linear ionic transport.
%In this work, we demonstrate that the mechanosensitive behavior of a graphitic nanochannel can be engineered through the deliberate design of surface charge patterns, which give rise to nonlinear ionic transport.
%In this study, we engineer mechanosensitivity by inducing a carefully placed surface-charge patch at one inlet of graphite-based ion channels. 
Our devices demonstrate a bipolar mechanosensitivity of the ionic conductance to pressure drops, which changes sign upon pressure drop reversal. The sensitivity reaches the ten millibar range on a wide range of salt concentration and nanochannel height. The detailed modeling of the electrohydrodynamic transport across the nanochannel allows rationalizing the pressure-dependent response of the conductance. This behavior is shown to be rooted in the \Rev{advection \QY{ of the electrical double layer (EDL) by water flow}  induced} 
%flow-modulated surface and diffusio-osmotic transport of a \Rev{concentration polarization} created
by the selective charging of one channel's inlet. We show that the \QY{ observed mechanosensitivity }is strongly enhanced by the low friction of water on graphite.
%With extensive theoretical modeling, we show that pressure sensing originates in the charge asymmetry combined with the high slip length of water on graphite.

This work \QY{ demonstrated a method for engineering artificial} nanofluidic channels, inspired by the precise arrangement of charged groups and binding patches, \Rev{instrumental to the selectivity of} biological porins \cite{murata2000structural,sui2001structural}. %\Rev{Such a device could be enclosed in a small deformable cell to serve as a bidirectional nanofluidic pressure sensor capable of detecting mbar-range pressure drops in-vivo without mechanical deformation and using a simple electrical readout.}
%Beyond the mere \LB{electrification patterns investigated} here, we envision a \LB{broad} potential in the functionalization of nanochannels inlets for advanced sensing and selectivity in nanofluidics.
\LBB{Integrated into a small deformable cell, the device could operate as a bidirectional nanofluidic pressure sensor, detecting millibar-range pressure drops through a simple electrical readout and without structural deformation.}

\begin{figure*}
	\centering
	\includegraphics[width=\linewidth]{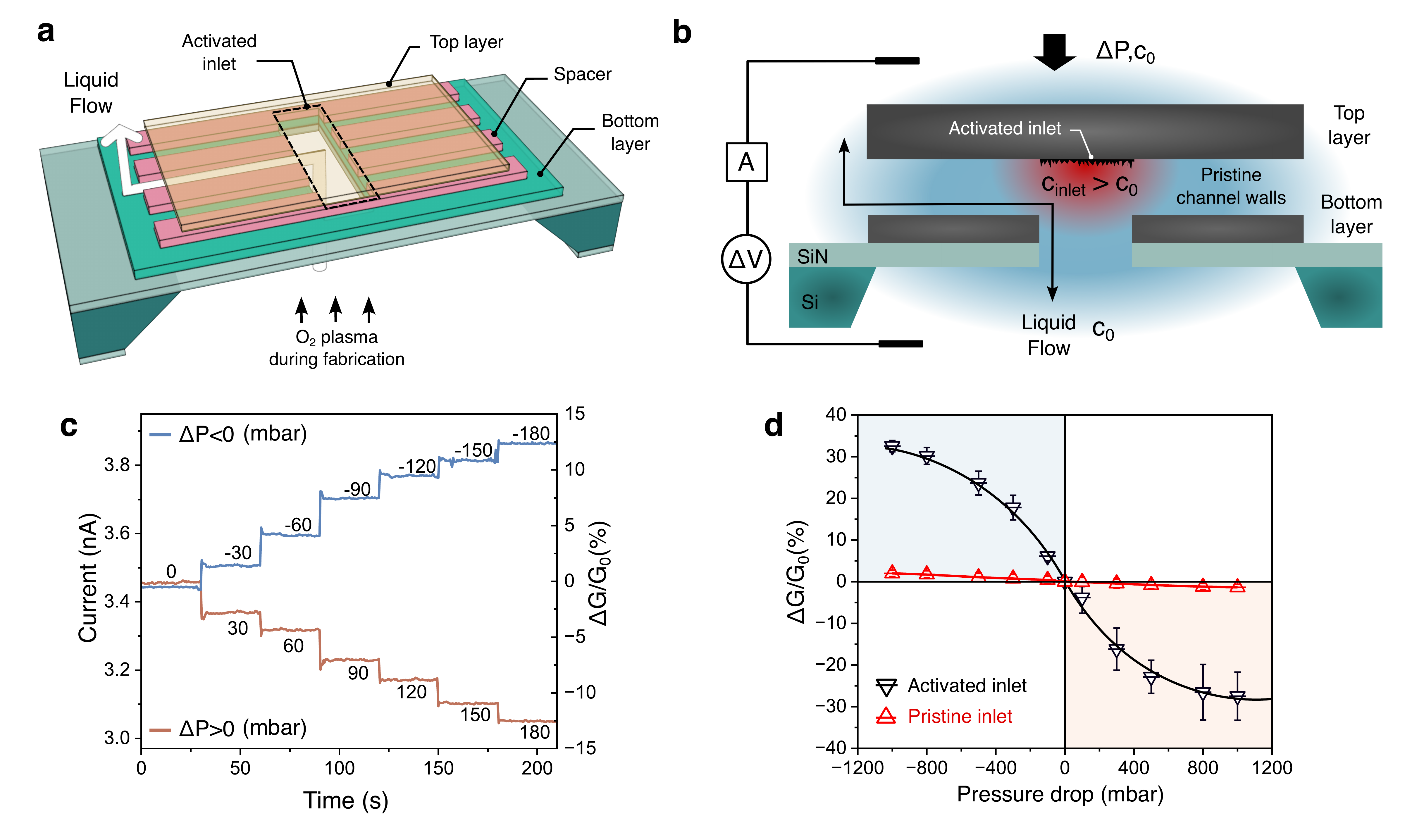}
	\caption{\textbf{{\color{black}Mechano-sensitive nanochannels }experimental \QY{setup}. }(\textbf{a-b}) A two-dimensional graphite nanochannel \Rev{fabricated by \QY{van der Waals} assembly} is placed between two reservoirs filled with aqueous KCl solution. A voltage drop is applied between two Ag/AgCl electrodes places in two reservoirs on either side of the channel. A pressure drop is also applied between the two reservoirs. By convention, the pressure drop is positive when the pressure is applied on the top side of the nanochannel, as indicated by the arrow on top. (\textbf{c}) The measured current {\color{black}from} a 1~nm activated channel in 1~M KCl solution under both positive and negative pressure drop steps \ML{and a voltage drop $\Delta V = 100$ mV.} (\textbf{d}) The relative conductance change $\Delta G/G_0$ as a function of applied pressure \ML{for both plasma-activated channels ($h =$ 1~nm, black symbols) and pristine channels ($h=$ 1.3~nm, red symbols) in 1~M KCl solution. Solid lines are guide for the eyes. }}
	\label{figure1_revision}
\end{figure*}

% \begin{figure*}
% 	\centering
% 	\includegraphics[width=\linewidth]{Revised_figs/Fig_solver_1.png}
% 	\caption{\textbf{{\color{black}Mechano-sensitive nanochannels }experimental set up. }(\textbf{a}) A two-dimensional graphite nanochannel is placed between two reservoirs filled with {\color{black}aqueous KCl solution}. A voltage drop is applied between two Ag/AgCl electrodes places in two reservoirs on either side of the channel. A pressure drop is also applied between the two reservoirs. By convention, the pressure drop is positive when the pressure is applied on the top side of the nanochannel, {\color{black} as indicated by the arrow on top}. (\textbf{b}) The measured current {\color{black}from} a 1~nm activated channel in 1~M KCl solution under both positive and negative pressure drop steps \ML{and a voltage drop $\Delta V = 100$ mV.} (\textbf{c}) The relative conductance change $\Delta G/G_0$ as a function of applied pressure \ML{for both plasma-activated channels ($h =$ 1~nm, black symbols) and pristine channels ($h=$ 1.3~nm, red symbols) in 1~M KCl solution. Solid lines are guide for the eyes. } %The inset is the zoomed-in version for the channel with pristine surfaces.
% 	(\textbf{d-g}) Concentration-dependent mechanosensitivity curves for {\color{black} channels of different }height $h \in \{1, 2.4, 4, 7\}$~nm for a set of pressures ranging between 0 and 1~bar. \ML{Note that the same dataset is replotted in Supplementary Figure 9 as a function of pressure instead of concentration.}}
% 	\label{figure1}
% \end{figure*}

\RevEditor{\section*{Results}}

\subsubsection*{Nanochannel fabrication}
Two-dimensional (2D) graphite slits are fabricated by van der Waals assembly following the technique introduced in Ref.\cite{radha2016molecular}, see Figure \ref{figure1_revision}\textbf{a} and details in Supplementary Methods. In brief, the channels were made of multilayer graphite spacers sandwiched between bottom and top graphite crystal. The sandwiched assembly sits on a suspended silicon nitride (SiN) membrane with rectangular opening, allowing ion transport \QY{ (the white arrow in Figure 1a indicates the ion transport pathway)}. The spacer layer was patterned using electron beam lithography (EBL) \QY{ and reactive ion etcher (RIE)} to form tens of channels in parallel, which act as the side walls of the channels and define their height.

At the end of the fabrication process, we exposed the \QY{ backside of the device} to a mild Ar/O$_2$ plasma \QY{ (as shown by the black arrows in Figure 1a)} in order to create a patch of "activated carbon" at the bottom-side inlet of the channels. \LB{Such an activation generates} a strong surface charge \LB{on the surface of graphite when immersed} in water  \LB{as compared to} pristine graphite \cite{emmerich2022enhanced}. As a control, we also produced "pristine-inlet" samples prepared in the exact same way to the exception of the activation step. We show in the following that this selective activation is instrumental to the pressure sensing capability of our channels.

\subsubsection*{Ion transport measurement}
We measured the ionic transport properties of the channels by placing the device in a fluidic cell between two tightly bound reservoirs which are filled with potassium chloride (KCl) aqueous solutions. A bias voltage $\Delta V$ and pressure drop $\Delta P$ can be applied across the nanochannel between the two reservoirs. \Rev{All experiments presented here were performed at room temperature.} In order to benchmark the experimental procedure, we first characterized the bare ($\Delta P=0$) conductance of the channels at different concentration, as well as the ionic current under voltage  and concentration drops, \QY{ respectively}; see Supplementary Note 1. These results show excellent agreement with the pre-existing body of experimental work on 2D graphite channels \cite{esfandiar2017size,mouterde2019molecular,keerthi2021water,robin2023disentangling}.

\LB{Turning now to the combined $\Delta V$-$\Delta P$ drivings, we report in} Figure \ref{figure1_revision}\textbf{c}  the time traces of ionic current measured as a function of pressure under a 0.1 V bias in a 1-nm-height channel. We observe a clear drop of current in steps mirroring the pressure steps at positive $\Delta P$ (pressure {\color{black}applied from} the top-side of the channels), and vice versa, an increase of currents at negative $\Delta P$, when pressure is applied from the back-side of the channels. The measured current-voltage \LB{response is linear} at all pressures -- see Supplementary Figure 4 -- suggesting that the system is characterized by a pressure dependent conductance $G(\Delta P)=I_e/\Delta V$ (with $I_e$ the measured ionic current under the voltage $\Delta V$ and pressure drop $\Delta P$).

\LB{Let us note that} throughout all measurements, the streaming current, induced by the advection of \Rev{EDLs} by the water flow and measured at zero voltage drop $I_{\rm str} = I(\Delta P,\Delta V=0)$, remains in the tens of picoAmps range (see Supplementary Figures 3 \& 6), while the channels' conductance is of a few tens of nS. Therefore, for any bias voltage above a few milliVolts, the streaming contribution is negligible.

In Figure \ref{figure1_revision}\textbf{d}, we plot the relative mechano-conductance $\Delta G/G_0$, where $G_0 = G(\Delta P=0)$ and $\Delta G = G(\Delta P)-G_0$ of \QY{1-nm-height channels} in 1~M KCl solution as a function of pressure $\Delta P$. We compare results from both activated-inlet channels (black symbols) and pristine-inlet channels (no plasma exposure, red symbols). Despite similar channel height ($h$), we observe strongly contrasting trends. The \LB{activated} device, \LB{which received a \QY{post plasma} treatment of the inlet}, shows a \LB{strong} sigmoid-shaped modulation \LB{of the conductance, with a relative increase bypassing} 30 percent, \LB{and} a saturation of the pressure modulation above 500~mbar. \LB{In contrast}, for channels with \LB{pristine, non-activated} inlets, a much smaller modulation -- less than 2 percent at 1~bar -- is observed. Notably, the sign of the pressure drop directly determines that of the conductance change. That is, conductance drops when pressure is applied on the top side of the chip, whereas it increases when pressure is applied on the back, \ML{yielding an odd response to pressure drops $(G(-\Delta P) = -G(\Delta P))$} which is \LB{rarely observed} in mechanosensitive ion channels.

\subsection*{Modelling of the bipolar pressure-dependent ion conduction}

%\LB{We now show that the mechano-sensitivity can be accounted in our geometry for by the non-linear couplings between ion conduction and fluid transport, which are boosted by the surface charge pattern in the present system. }
We now demonstrate that the observed mechano-sensitivity in our geometry can be explained by the nonlinear couplings between ionic conduction and fluid transport, which are enhanced by the surface charge pattern characteristic of the present system.

However, before entering into the ion dynamics, let us first show that mechanical deformations can be ruled out as origins to the pressure-sensing behavior of our channels.
%\paragraph*{Ruling out mechanical deformations}
Such a mechanism was \QY{invoked} as a source of mechanosensitivity in some artificial solid-state systems \cite{davis2020pressure,macha2022stress} and we first estimate
%To explain the pressure-dependence of our channel's ionic conductance, we first carefully considered 
the mechanical deformations of the \QY{ nanochannel walls} and the SiN membrane. 
%since \LB{such a mechanism was invoqued as a source} of mechanosensitivity in some artificial solid-state systems \cite{davis2020pressure,macha2022stress}. 
Under a pressure drop $\Delta P$, \LB{the sagging of} the graphite \LB{top layer forming the nanochannel} (with thickness $H =$ 50~nm, channel's width $w=200$~nm and Young's Modulus $E=1$~TPa) is \LB{predicted as} $\delta = 5 \Delta P w^4 /32 E H^3$ \cite{yang2020capillary,fumagalli2018anomalously}. For $\Delta P$ as high as 1 bar, we find \LB{a deformation of} $\delta \sim 0.2$~pm, ruling out any measurable \LB{effect of the deformation of the channel wall on the conductance}. 
\LB{We also ruled out the deformation of the SiN support membrane as a source of pressure dependent conductance. To check this effect,} 
%The nanochannels are assembled on top of 500~nm SiN membrane, with 25 $\mu$m by 3 $\mu$m openings which can deform up to roughly $400$~nm under 1~bar pressure difference, according to finite-element calculation. To rule out conductance change sourced from the deformation of SiN window, 
we fabricated a 2.4-nm-\QY{height} channels on a much thinner SiN membrane (100~nm instead of 500 nm). We show in Supplementary Figure 5 that the \QY{{mechanosensitivity}} in these devices are similar for the 100~nm and 500~nm thick SiN membranes  despite a 125-fold difference in the stiffness of the SiN membrane. 
%Overall, we can safely rule out mechanical deformations as origins to the pressure-sensing behavior of our channels. %and S6\textbf{b}

%\paragraph*{Coupled flow and ion transport}
\subsubsection*{The activated inlet as a source of disposable charge carriers : \Rev{a 1D advection model}}
The absence of mechanical effect thus suggest an electrohydrodynamic origin for the \QY{{mechanosensitivity}}. The mechano-sensitive response observed here is however singular as compared to some previous observations.
%REVOIR
%Pressure gating of ion transport that is not caused by the deformation of nanochannels or nanopores has also been reported before. In strongly rectifying glass nanopipettes for instance, pressure was shown to yield strongly non-linear conductance gating by deforming the voltage-induced space charge zone \cite{jubin2018dramatic}. In graphene nanopores, a similar model based on the advection of potential-induced charges by the flow, was used to explain a symmetric and linear G(P) law \cite{jiang2022nonlinear,paul2024nanoscale}. 
%\LB{NO COMPRENDO}
First, the pressure response reported here is odd with respect to $\Delta P$: $\Delta G(-\Delta P) = -\Delta G(\Delta P)$, in contrast to the symmetric dependence in  carbon nanotubes and across graphene nanopores reported before \cite{marcotte2020mechanically,jiang2022nonlinear}. Furthermore, we invariably measured perfectly linear current-bias response at all pressure drops (see Supplementary Figure 4), and this rules out mechanisms that are intrinsically non-linear in $\Delta V$ \cite{jubin2018dramatic,jiang2022nonlinear,paul2024nanoscale}. This implies that the present phenomenon originates from a distinct set of underlying physical mechanisms.

%The geometry of the system is sketched in Figure \ref{figure1_revision}\textbf{a-b}.
A crucial observation in our experiments is the role of the (oxygen-plasma) activation of the channel inlet, which considerably amplifies the pressure response of the conductance. The activation induces a strong surface charging of the \QY{graphitic} surface in water, typically of the order of several hundreds mC/m$^2$ \cite{emmerich2022enhanced}, \QY{serving as a high surface charge patch at the bottom }inlet of the channel (Figure \ref{figure1_revision}\textbf{a-b}), while the \QY{nanochannels} remain pristine. This yields a strong surface charge asymmetry between the two inlets of the nanochannel, as well as \QY{between bottom inlet and the interior} of the channel.

\begin{figure*}
	\centering
	\includegraphics[width=\linewidth]{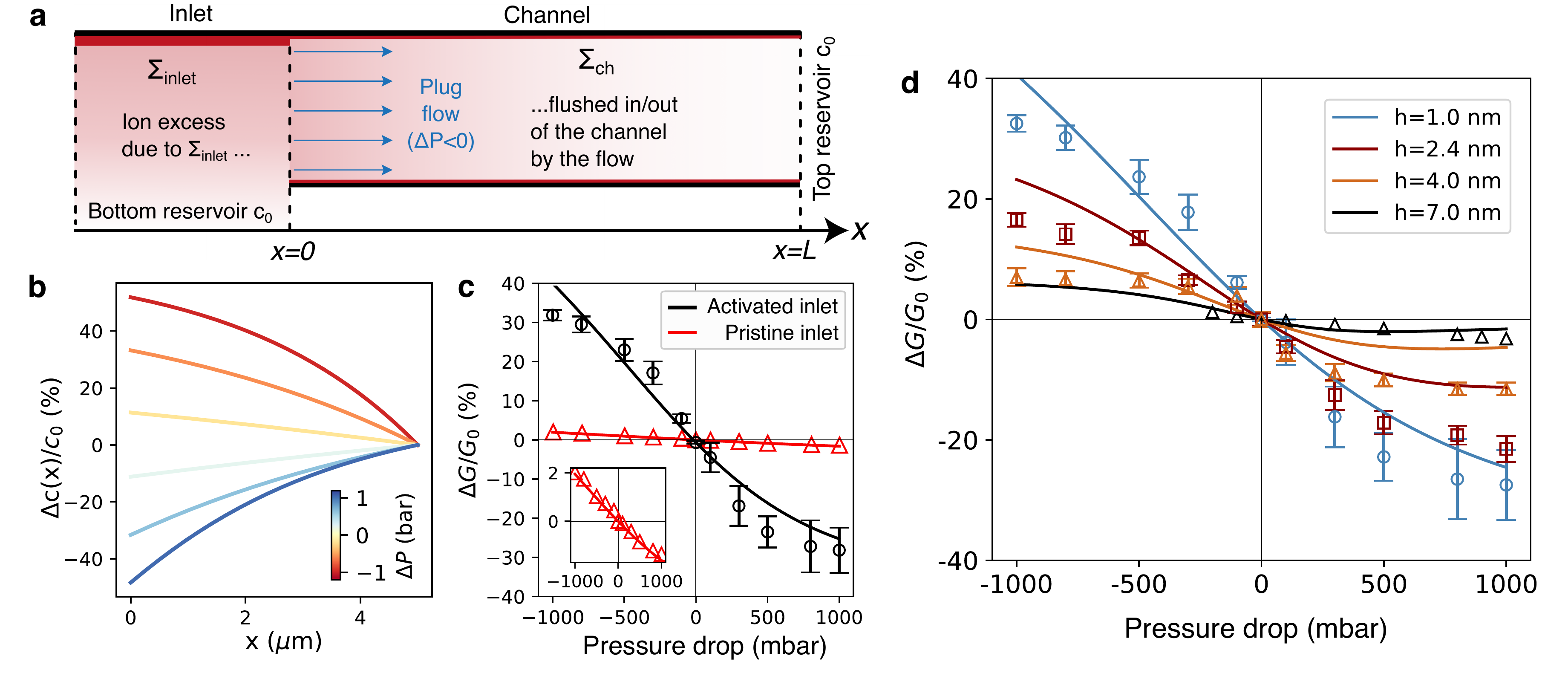} 
	\caption{\textbf{\LB{An \QY{electrohydrodynamic} model for \QY{mechano-sensitivity}}}(\textbf{a}) Sketch of the model: counter charges {\color{black}near} the activated inlet are advected by the water flow {\color{black}with applied pressure}, resulting in a pressure-dependent concentration profile in the channel. \QY{Darker color represents higher concentration} \Rev{(\textbf{b}) Theoretical predictions for the pressure-dependent relative concentration drop profiles in a \QY{1-nm-height} channel plotted for $h=1$~nm, $L=5$~$\mu$m, $b=20$~nm, $\Sigma_{\rm ch} = -25$~mC/m$^2$ and $\Sigma_{\rm inlet}=-150$~mC/m$^2$ at $c_0 = 1$~M, closely matching experimental conditions.}  (\textbf{c}) \LBB{Experimental results for the} mechano-conductance of \QY{1-nm-height} channels with respectively a pristine (red) and an activated inlet (black), versus the theoretical predictions: Solid lines are 1-parameter fits with the model, with \Rev{the slip length} $b$ fixed at 20~nm, \Rev{the channel surface charge density} $\Sigma_{\rm ch} = -25$~mC/m$^2$ and \Rev{the inlet's surface charge density} $\Sigma_{\rm inlet}$ as a single free parameter. \Rev{From the fit of the activated-inlet (black curve) and pristine-inlet (red) data, we obtain respectively $\Sigma_{\rm inlet}=-204\pm10$~mC/m$^2$ \QY{for activated channels} and $\Sigma_{\rm inlet}=-50\pm1$~mC/m$^2$ \QY{for pristine-inlet channels}.} (\textbf{d}) Mechano-conductance measurements at $c_0$=1~M on the whole pressure range between -1~bar and 1~bar for $h\in \{1,2.4,4,7\}$~nm \LBB{against the theoretical predictions.} \LBB{Fit parameters are, respectively,} \Rev{$\Sigma_{\rm inlet} \in \{-204,-160,-127,-102\}$ $\pm 10$} ~mC/m$^2$,  $b=$ 20~nm and $\Sigma_{\rm ch} = -25$~mC/m$^2$.}
	\label{figure3}
\end{figure*}

%\subsection*{The inlet-charge as a source of disposable charge carriers}
\paragraph*{General mechanism and transport framework --}
Starting from the Poisson-Nernst-Planck-Stokes (PNPS) framework, the following mechanism emerges from the solution of the transport equations: 
\LBB{the ions in the EDL close to the activated surface are flushed towards (or away from) the nanochannel but
entering the nanochannels these charges face a discontinuity of surface diffusivity. This leads to an accumulation (or depletion) of ions at the entrance,}
%the charged patch at the entrance of the channel \Rev{creates a region of enhanced ion concentration,} 
which is advected under the pressure-driven flow (see Figure \ref{figure3}\textbf{a}). This results in a pressure-dependent \Rev{ion-concentration profile} in the nanochannel. The ionic conductance, which is proportional to the number of free ions in the nanochannels, will increase \Rev{for negative pressure drops which flush the inlet concentration polarization into the channel (as sketched on Figure \ref{figure3}\textbf{a}) and decrease for positive pressure drops which flush the concentration polarization out of the nanochannel.}

%profile in turn controls the electrical conductance (conductivity $\sigma(x)\propto c(x)$ from the Drude model). We model the activated inlet as a first channel with surface charge $\Sigma_{\rm inlet}/2$ in series with the pristine graphite channel of surface charge $\Sigma_{\rm ch}$ (see schematic in Supplementary Figure 9). This model relies on the robust linear response to bias we measured at all pressures and concentrations (see Supplementary Figure 4), allowing us to treat the voltage drop as a perturbation and thus to simply consider the pressure-dependent concentration profile.

\paragraph*{Capillary pore model and scale separation --}
While the scenario is relatively simple, the detailed mechanism \Rev{combines} diffusion, convection, surface conduction as well as diffusio-osmotic transport, making the implementation of the model quite involved. However it can be simplified to obtain analytical predictions for the ion conduction amenable for direct comparison with the experiments.
%The geometry is described in Figure \ref{figure3}\textbf{a}.  Flow is expected to advect the charges within the nanometric Electric double layer (EDL) on the activated surface of the reservoir and mainly from the vicinity of the nano-channels entrance.

\Rev{To simplify} the calculations while keeping the main physical ingredients, we reduced the geometry to that of a continuous slit, with height $h$, connecting \Rev{the nanochannel and the \QY{activated} inlet region, with respective surface charges $\Sigma_{\rm ch}$ and $\Sigma_{\rm inlet}$; see Supplementary Figure 10 and Supplementary Note 2.} %We anticipate from the calculations that the length ratio $L_1/L$ is mostly irrelevant and does not change the physical conclusions. We will accordingly choose $L_1=L$ in the following. %(since varying the ratio $L_1/L$ is mostly irrelevant and does not change the physical conclusions).
Overall, this simplified geometry retains the charge heterogeneity and the feeding of the nanochannel by the charges within the \Rev{EDL at the inlet's surface} as the main ingredient for the observed behavior. We further validated this \QY{simplified} assumption by comparing our analytical predictions to exhaustive finite element calculations \Rev{in a geometry which very closely reproduces experimental conditions. All details are given in Supplementary Note 3}. % , see below.

The description then follows the capillary pore (or space charge) model %introduced by Osterle {\it et al.}, and
discussed in \cite{peters2016analysis}. \Rev{Under a pressure drop $\Delta P$, the averaged water velocity in the channel writes} $ v_w = \Delta P \frac{h^2}{12\eta L}\Bigl(1+6 b/h\Bigr)$ \Rev{where $h$ and $L$ are channel's height and length, $b$ the slip length, $\eta$ is water's dynamic viscosity}. We introduce the Péclet number $\Pe = v_wL/D$ comparing the advection at scale $L$ to diffusion (\Rev{with $D$  the diffusion coefficient}):
\begin{equation}
    \Pe = \frac{hb\Delta P}{2\eta D}
\end{equation}
where we assume $b\gg h$ in the nanochannel, an assumption which will be validated below. Turning now to ionic transport, \Rev{we focus on the limit where the Debye length is smaller than the channel height, which is relevant in our experimental conditions.} We accordingly use the scale-separation methodology introduced by Poggioli et al. \cite{poggioli2019} to rewrite the ionic transport equations by integrating transport over the EDL contributions. Details are given in Supplementary Note 2.
% and compute the concentration profile \LB{across the heterogeneous channel.} 
%The key idea here is to separate lengthscales and split the problem along the channel height direction z ($h\sim 1-10$~nm $\sim\lambda_D$) and along channel length direction x  ($L\sim 5 \mu$m $\gg \lambda_D$). One then solves the equilibrium Poisson Boltzmann problem along z, taking into account the charge separation and variations in electric potential. 
%Along x however, equilibrium is broken by the flow but we assume electroneutrality to be maintained. 

\Rev{Taking into account the diffusion, surface contributions, and convection, the ion flux is calculated from the salt-concentration profile $c(x)$} as
\begin{equation}
J = -{\cal A}D  (1+\mathcal{F}[\Du(x)]) \left( \partial_x c(x) -\Pe\, \frac{c(x)}{L}\right)
\label{eqn:Jsol_genera2l}
\end{equation}
\Rev{where $\cal A$ is the channel's cross section and} we introduced \Rev{the local Dukhin number $\Du(x)={ \Sigma(x)/h c(x)}$ accounting for the ratio of surface charge $\Sigma$ to bulk ion content $hc(x)$.} The term $\mathcal{F}[\Du(x)]$ accounts for the surface contribution to transport within the EDL and is calculated as
%$\mathcal{F}$ is a dimensionless function of the Debye length $\lambda_D$, the slip length $b$ and the Dukhin number $\Du = \Sigma/ch$ which compares surface to bulk conduction in a channel of height $h$ and surface charge $\Sigma$. $\mathcal{F}$ is 
%a measure of the number of additional free ions in the channel due to the surface charge,  writing:
\begin{equation}
	\mathcal{F} [\Du]= 2\frac{\lambda_D}{h} \Bigl( \sqrt{1+\Bigl[ \frac{h \Du}{2\lambda_D} \Bigr]^2} - 1 \Bigr)
\end{equation} \Rev{where $\lambda_D = \sqrt{\epsilon k_{\rm B}T/2e^2 c}$ is the Debye length.}

Eq.(\ref{eqn:Jsol_genera2l}) can be solved for the concentration profile $c(x)$ inside the nanochannel and in the \Rev{inlet region}.  As boundary conditions, $c(x)$ asymptotically matches the bulk ion concentration $c_0$ \Rev{in both reservoirs at $x \to \pm \infty$}. Furthermore, we match concentration profiles and ion fluxes between the inlet region and channel. \LBB{Results are analytical for low P\'eclet. For high P\'eclet, }\Rev{the one-dimensional differential equation is then \QY{solved} numerically to obtain the \QY{concentration} profile $c(x)$ and ion flux $J$ in the nanochannel.}
\Rev{Finally, from the salt concentration profile,} we compute the conductance \Rev{from the local Ohm's law}
\begin{equation}\label{eq:local_Ohm}
	G \sim 1/\int_{0}^{L} \Bigl[c(x)+\Sigma_{\rm ch}/h\Bigr]^{-1} dx
\end{equation} and the relative mechanosensitivity $\Delta G/G_0$ which directly compares to experiments.

\Rev{\subsubsection*{Diffusio-osmotic correction}}

Now under the flow-induced concentration gradient, a diffusio-osmotic (DO) flow \Rev{builds up and adds} to the bare pressure-driven flow, \RevEditor{echoing experimental findings in carbon nanotubes \cite{Lokesh2018}}. The DO velocity takes the form
$v_{\rm DO}= K_{\rm DO} \Delta c\Rev{(x=0)}/L$
where the DO mobility can be estimated as \cite{marbach2019}
\beq
K_{\rm DO} \simeq {k_BT\over \eta} \lambda_D^2\left(1+{b\over \lambda_D}\right)
 \approx {k_BT\over \eta} \lambda_D b
\eeq
with \Rev{$k_{\rm B}T$ the thermal energy, $\lambda_D$ the Debye length and \QY{the slip length} $b\gg \lambda_D$ }. \Rev{To compute $v_{\rm DO}$, we estimate the concentration drop across the channel at first order in $\Delta P$.}

%\paragraph*{Concentration gradient at low $\Delta P$ --}
\Rev{At low pressure drops ($\Pe\ll1$),} one may deduce analytically the excess concentration at the entrance of the channel as
\begin{equation}
\frac{ \Delta c (x=0)}{c_0}=\Gamma \times \Pe
\end{equation}
with $\Gamma$ being a function of the Dukhin numbers associated with the channel \Rev{($\Du\ch$)} and activated patch \Rev{($\Du\p$)} surface charge:
\begin{equation}
\Gamma = \frac{ \mathcal{F}[\Du\p]-\mathcal{F}[\Du\ch] }{2+\mathcal{F}[\Du\p]+\mathcal{F}[\Du\ch]} %term + \frac{L}{L_1}(1+\mathcal{F}[\Du\p])
\label{Gamma}
\end{equation}

%In this low pressure-drop limit, the flow-induced concentration profile in the nanochannel is mostly linear. 

%\paragraph*{\Rev{Diffusio-osmotic boost} --}

Now, since the concentration drop is induced by the flow itself $\Delta c \Rev{(x=0)}=c_0\,\Gamma\, \Pe\propto v_w$, one deduces that the DO velocity is proportional to the pressure-driven
water velocity, in the form
%$v_{\rm DO}= c_0{k_BT\over \eta D}\lambda_D^2\left(1+{b\over \lambda_D}\right)   \times \Gamma \times {v_w }$. 
$v_{\rm DO}= c_0{k_BT\over \eta D}\lambda_D\,b   \times \Gamma \times {v_w }$. 
The total water velocity is now 
%$v_{\rm eff}= v_w+v_{\rm DO}=v_w\times \left( 1 + c_0{k_BT\over \eta D}\lambda_D^2\left(1+{b\over \lambda_D}\right)   \times \Gamma \right)$.
$v_{\rm eff}= v_w+v_{\rm DO}=v_w\times \left( 1 + c_0{k_BT\over \eta D}\lambda_D\, {b}  \, \Gamma \right)$.
Accordingly  DO transport renormalizes the Péclet number as
%\beq 
%\Pe_{\rm eff} = \Pe_0 \left( 1 + c_0{k_BT\over \eta D}\lambda b  \times \Gamma \right)
%\eeq
\begin{equation}
%	\Pe_{\rm eff} = \Pe \Bigl[ 1+\frac{\ML{c_0} k_B T}{\eta D}\lambda_D^2 (1 + \frac{b}{\lambda_D}) \Gamma\Bigr]
	\Pe_{\rm eff} = \Pe \Bigl[ 1+\ML{c_0}\frac{ k_B T}{\eta D}\lambda_D \, {b}\, \Gamma\Bigr]
	\label{Peeff}
\end{equation}
where the coefficient $\Gamma$ is defined in Eq.(\ref{Gamma}). Due to slippage effect, this correction to the Péclet number is large and hydrodynamic slip acts as a booster of the mechanosensitive effect.

% The concentration drop at the entrance of the nanochannel \Rev{thus writes:
% \begin{equation}
% \frac{ \Delta c(x=0)}{c_0}=\Gamma \times \Pe_{\rm eff}
% \end{equation}}
% with the effective Péclet number defined in Eq.(\ref{Peeff}).

\vspace{1em}

\subsubsection*{\Rev{Low pressure-drop limit}}

\Rev{In the low pressure-drop limit}, the relative mechano-conductance $\Delta G/G_0$ takes the final expression
\begin{equation}
	\frac{\Delta G}{G_0} = \frac{1}{1+\Du_{\rm ch}}\cdot \frac{\mathcal{F}\left[{\rm Du_{inlet}}\right]-\mathcal{F}\left[{\rm Du_{ch}}\right]}{2+\mathcal{F}\left[{\rm Du_{inlet}}\right]+\mathcal{F}\left[{\rm Du_{ch}}\right]}\times  \frac{\Pe_{\rm eff}}{2} 
\label{DGfinal}
\end{equation}
\Rev{which scales linearly in the pressure drop $\Delta P$ (see Supplementary Note 2). Interestingly, the response} is shown to vanish \Rev{in the absence of charge contrast} (when $\Sigma_{\rm inlet}=\Sigma_{\rm ch}$). This result therefore highlights the key role of the charge pattern in the mechano-sensitive response, in agreement with the experimental observations, see Figure \ref{figure1_revision}\textbf{d}. 
%\Rev{To further explore the model's response and taking advantage of this relatively simple low-$\Delta P$ formula, it is especially interesting to consider the impact of the bulk salt concentration $c_0$ -- which can be easily varied -- on mechanosensitivity. A first guess may be that as a surface effect, the \textit{relative} conductance modulation, $\Delta G/G_0$ should be maximum at low concentration. We will now see that it is not really that simple:}
\LBB{The prediction in Eq.(\ref{DGfinal}) shows a non-trivial dependence of mechanosensitivity on the bulk salt concentration $c_0$, as we now discuss.}

%As \Rev{we will now show}, the predictions from Eq.(\ref{DGfinal}) explain the trends.%are in agreement with all experimental observations in various regimes.

\paragraph*{High concentration regime --}
\LBB{First,} at high salt concentration, surface contributions become negligible as the Dukhin numbers vanish \Rev{and} $ \mathcal{F}\left[{\rm Du}\right] \simeq \frac{h}{4 \lambda_D} \Du^2\rightarrow 0$. Further assuming $\Sigma_{\rm inlet}\gg \Sigma_{\rm ch}$ due to plasma activation, the Péclet-dependence of the conductance then reduces to the expression
\begin{equation}\label{eq:linearPe_highc}
	\frac{\Delta G}{G_0} \approx   \frac{1}{16h \lambda_D c_0^2}\frac{\Sigma_{\rm inlet}^2}{4}\times \Pe_{\rm eff}
\end{equation} The relative mechano-conductance $\Delta G/G_0$ scales as $1/(\lambda_{\rm D}c_0^2) \sim c_0^{-3/2}$, dropping at high concentration. \Rev{Interestingly, if the surface charge is regulated by salt concentration with a power $\alpha$ ($\Sigma_{\rm inlet} \sim c_0^{\alpha}$) as previously reported on carbon surfaces \cite{secchi2016scaling,emmerich2022enhanced}, the decay at high-concentration is  weaker, scaling as $\sim c_0^{-3/2+2\alpha}$.}

\paragraph*{Low concentration regime --}%Eq 92 of the theoretical notes
At low salt concentration \Rev{on the other hand}, the Dukhin number is large, so that $\mathcal{F}\left[{\rm Du}\right] \simeq  \Du $. Again assuming that $\Sigma_{\rm inlet}\gg \Sigma_{\rm ch}$, we find:
\begin{equation}\label{eq:linearPe_lowc}
	\frac{\Delta G}{G_0} \approx \frac{1}{2 \Du_{\rm ch}}\times \Pe_{\rm eff} 
\end{equation}

Eq.(\ref{eq:linearPe_lowc}) predicts that mechanosensitivity vanishes linearly at low salt concentration $c_0$. \LBB{
This results from the dominating surface conductance in the channel at high Dukhin.} 
%\Rev{as the channel's EDL, which is unaffected by the inlet-EDL advection, dominates conductance.} %\Rev{a trend that we will later show to agree with experimental results.}% It therefore reproduces the counterintuitive \LB{experimental observation that the mechano-sensitive} effect vanishes as the salt concentration \Rev{drops to zero (see Figure \ref{figure2_revision}\textbf{a-d} and Supplementary Figure 9).}

\LBB{Altogether, this analysis suggests that mechanosensitivity vanishes both at high and low concentration, thus}
%\Rev{From this limiting-case analysis, we anticipate that somehow counter-intuitively, our EDL-advection based mechano-sensitivity vanishes not only at high but also at low concentration, 
\Rev{resulting in a non-monotonous variation with concentration. We will test this prediction against experimental results in the following section.} %\Rev{For a constant inlet surface charge ($\alpha=0$), we find that indeed, $\Delta G/G_0$ is very strongly suppressed at high concentration (see Supplementary Figure 11). We show on Figure \ref{figure2_revision} that the concentration regulation of $\Sigma_{\rm inlet}$ strongly dampens this supression, } explaining the observed drop in magnitude of $\Delta G/G_0$ above 1~M across pressures, as well as with confinement, see Figure \ref{figure2_revision}\textbf{a-d}. 

\subsubsection*{\Rev{General solution}}

\paragraph*{\Rev{Concentration profile --}}

\LBB{Beyond the low-\Pe~limit, the EDL-advection model in Eq.(\ref{eqn:Jsol_genera2l})  is solved }
%\Rev{To use the EDL-advection model beyond the low-\Pe~limit, we solve Eq.(\ref{eqn:Jsol_genera2l}) 
\Rev{numerically to obtain the concentration profiles $c(x)$ inside the nanochannel and in the inlet region (see Supplementary Note 2 and Figure \ref{figure3}\textbf{b}). On Figure \ref{figure3}\textbf{b}, we plot the differential concentration profiles $\Delta c(x)/c_0$ for a given set of parameters which closely match the experimental conditions (see below).}
\LBB{The profile shapes illustrate the diffusion-advection balance under increasing flushing effects at high P\'eclet.}
%We find that the competition of salt advection from the activated inlet with its diffusive escape from the pristine inlet results, at the steady state, in a curved concentration profile along the channel's length.} %All details on the calculations are given in Supplementary Note 2.

\paragraph*{\Rev{Mechanosensitivity --}}

We report the \Rev{relative \QY{mechano-conductance} $\Delta G/G_0$ -- computed from the $c(x)$ profile with the local Ohm's law (Eq.(\ref{eq:local_Ohm})) -- in} Supplementary Figure 11 for $h =$ 1~nm, $L=5$~$\mu$m, $\Sigma_{\rm ch}=-25$~mC/m$^2$ and $\Sigma_{\rm inlet}=-150$~mC/m$^2$ for a range of slip length and salt concentrations. $\Delta G/G_0$ is found to be an odd function of pressure, with a magnitude in the order of a few tens of percent for $b\sim 20$ nm and is linear in $\Delta P$ at low Péclet  number. The results also highlight a saturation at high $\Delta P$, \Rev{which we find to be slightly asymmetric  -- stronger for positive pressure drops.} %in good agreement with the experimental results of Figure \ref{figure2_revision}.
\Rev{In Supplementary Figure 11, we also plot the prediction for $\Delta G/G_0$ \textit{versus} $c_0$ at a fixed slip length $b=20$~nm for a range of slit heights. As anticipated, we find that $\vert \Delta G/G_0 \vert$ is suppressed at both high and low concentrations. Importantly, the optimum concentration $c_0^{\rm max}$, at which $\vert \Delta G/G_0 \vert$ is maximum, shifts downwards with increasing slit height, from roughly 1~M at $h=1$~nm to 200~mM at $h=4$~nm.}

\begin{figure*}
	\centering
	\includegraphics[width=0.8\linewidth]{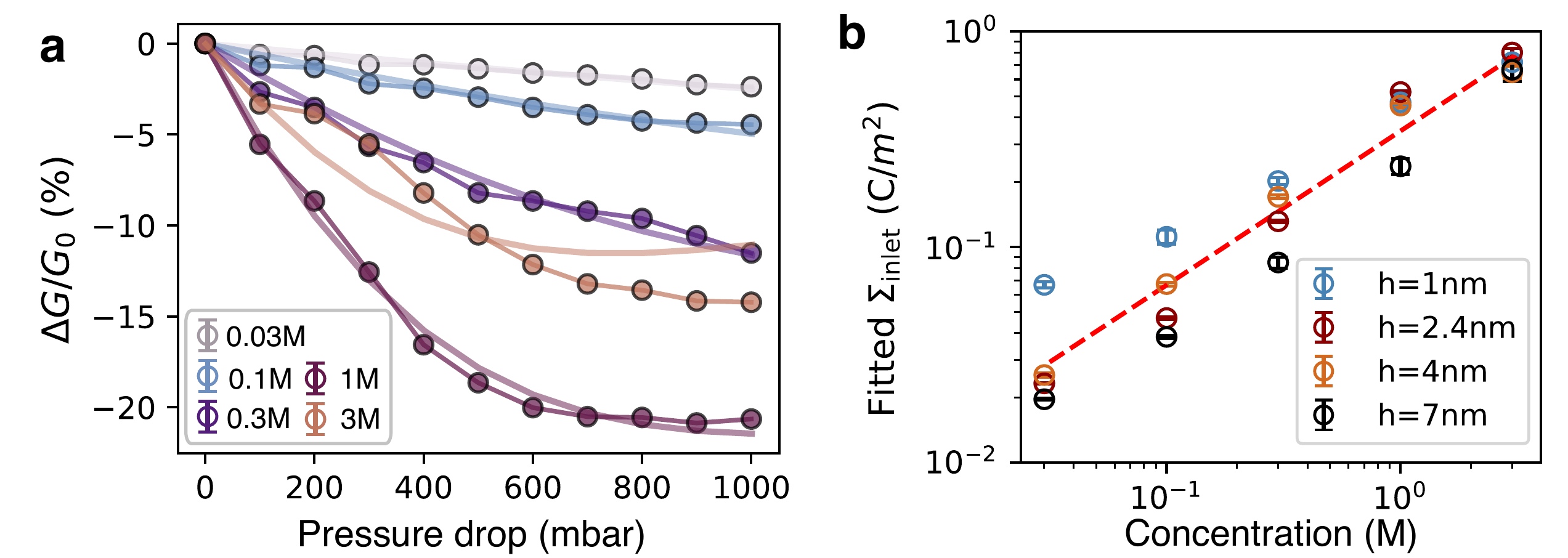} 
	\caption{\textbf{Concentration dependency of the mechano-sensitive effect and fits with the theoretical model} (\textbf{a}) Mechano-conductance \textit{versus} pressure curves for a channel with height $h=$ 2.4~nm for $\Delta P$ ranging between 0 and 1~bar. Solid lines are \ML{a 1-parameter fit with the theoretical} model, assuming a slip length of 20 nm, a pristine graphite surface charge of -25 mC/m$^2$ at $c_0$=1~M and a regulation of $\Sigma_{\rm ch} \sim c_0^{1/3}$ consistent with existing measurements on carbon nanotubes \cite{secchi2016scaling}. (\textbf{b}) \LB{Surface charge} $\Sigma_{\rm inlet}$ \LB{for each ion concentration} extracted from the fits \LB{of the theoretical predictions} for various channel height (\ML{see Supplementary Figure 9}). We find that the data \LB{for all confinements} collapse on a power-law dependence, $\Sigma_{\rm inlet} \sim c_0^{\alpha}$, \Rev{with $\alpha=0.71 \pm 0.05$ and independent of channel height, see the red dashed line.}}
	\label{figure4}
\end{figure*}

\subsubsection*{Finite element calculations}
\Rev{To support the predictions of the 1D EDL-advection model, we} performed finite element calculations of transport in the original geometry of the system, accounting \Rev{precisely for the converging flow at the inlet-nanochannel interface and for charge separation regions}. \Rev{Using Comsol, we} solved accordingly the complete set of non-linear \Rev{Poisson-Nernst-Planck-Stokes (PNPS)} equations for the 2D geometry under consideration and \Rev{explored} the pressure dependence of the concentration profile, as well as the change in conductance. All details are provided in Supplementaty Note 3. Solving this finite element model numerically, we find a pressure-dependent concentration profile in the channel, consistent with findings from the \Rev{EDL-advection model (see Supplementary Figures 13-16)}. 	

\Rev{Using the finite element calculations, we checked that electric fields along $x$ decays over less than 1 nanometer and that neglecting charge-separation in the analytical model (Eq. (\ref{eqn:Jsol_genera2l})) has little effect on the resulting conductance (see Supplementary Figure 13-14).} The numerical results fully reproduce \Rev{the odd pressure response at low $\Delta P$, the asymmetric saturation at high $\Delta P$ and the order of magnitude of the $\Delta G/G_0$ (see Supplementary Figure 15). We also recovers the scaling $\Delta G/G_0 \sim \Sigma_{\rm inlet}^2$ predicted in the low-Dukhin number limit}, see Eq.(\ref{eq:linearPe_highc}) (see Supplementary Figure 16). Overall the finite element calculations \Rev{support the hypothesis made and  confirm the predictions of the 1D EDL-advection model.} \LBB{We now use the predictions to fit the experimental data.}

\RevEditor{\section*{Discussion}}
We discussed above that the theoretical predictions successfully reproduce the overall experimental observations and trends: an odd mechano-sensitive effect for the conductance and the saturation at large pressure. %, as well as the qualitative dependence on salt concentration.
\Rev{To fully characterize the pressure-sensing mechanism, we} further examined experimentally the respective roles of confinement and salt concentration on mechanosensitivity. We accordingly investigated the properties of graphite channels with various heights $h \in \{1, 2.4, 4, 7\}$ nm, which were all subject to inlet activation by plasma treatment following the procedure described above.

Let us now compare systematically the predictions of the model with the experimental results. The model depends on several surface properties which are difficult to estimate independently : the surface charge of the channel and inlet and the slip length \Rev{of the channel walls. }% -- and their salt concentration dependence. %The fitting procedure is accordingly slightly under-constrained. and only a semi-quantitative comparison is expected. 
To account for our measurements, we fix the surface charge $\Sigma_{\rm ch}$ and slip length $b$ at the channel walls for all channels and salt concentrations. The slip length is fixed at $b=20$~nm to account for $G(\Delta P)$ measurements under imposed concentration drops in the pristine-inlet channels (see Supplementary Note 1, E). The value is also close to direct measurements for pure water on graphite (8 nm) \cite{maali2008measurement}. The wall's surface charge on the other hand, is fixed at $\Sigma_{\rm ch}=-25$~mC/m$^2$ for a 1~M KCl solution and assumed to scale as $c_0^{1/3}$ with salt concentration, \Rev{a regulation behavior reported }in previous work \cite{secchi2016scaling,esfandiar2017size}. As a consequence, we leave the inlet charge $\Sigma_{\rm inlet}$ as the only free parameter in the subsequent analysis. %This allows us to explain quantitatively almost all our results.

\paragraph*{Comparison of pristine and activated inlets --}
The predictions for the pristine and activated inlets are confronted in Figure \ref{figure3}\textbf{c} to the experimental results at 1~M \Rev{-- the solid line is a fit} \LBB{taking $\Sigma_{\rm inlet}$ as the only fitting parameter}. The theoretical prediction successfully describes the pressure variation, and the difference in magnitude of the mechano-sensitive effect is accounted for by using different surface charge on the inlet surface. Fitting the results under the assumption that $b=$ 20~nm and $\Sigma_{\rm ch}=-25$~mC/m$^2$, we obtained values of \Rev{$\Sigma_{\rm inlet}=-50 \pm 1$~mC/m$^2$ and  $-204$ $\pm$ 10~mC/m$^2$} for the surface charge on the pristine-inlet and activated-inlet channels respectively. \Rev{Note that the concentration profiles in Figure \ref{figure3}\textbf{b} were obtained with the parameters fitted to the activated-inlet channel data.}

\paragraph*{Influence of channel height --}
The influence of the channel height is reported \Rev{in} Figure \ref{figure3}\textbf{d}, where we plot the pressure-dependent mechano-conductance measured at 1~M concentration in channels of various heights h$\in \{1,2.4,4,7\}$~nm (all are activated at the inlet).  Here, we fit our model to the experimental data with the inlet charge as the only free parameter (here again $b=$ 20~nm and $\Sigma_{\rm ch}=-25$~mC/cm$^2$). Our model is shown to reproduce the experimentally observed mechano-conductance dependence for all confinement size $h$, with inlet charges changing in a relatively small interval, between \Rev{$-100$ and $-200$~mC/m$^2$}. In particular, it accounts quantitatively for the sigmoid-shaped relative mechano-conductance and its suppression as the channel's height increases, reducing from $\sim 30$ percent at 1~nm down to a few percent at 7~nm.
%The complementary experiments under \Rev{an imposed} concentration drop $\Delta c_0$ at the boundaries further strengthen the model, which accounts for all result without any free parameter with a slip length of 20 nm, see Supplementary Note 1 E and Supplementary Figure 8.

\begin{figure*}
	\centering
	\includegraphics[width=\linewidth]{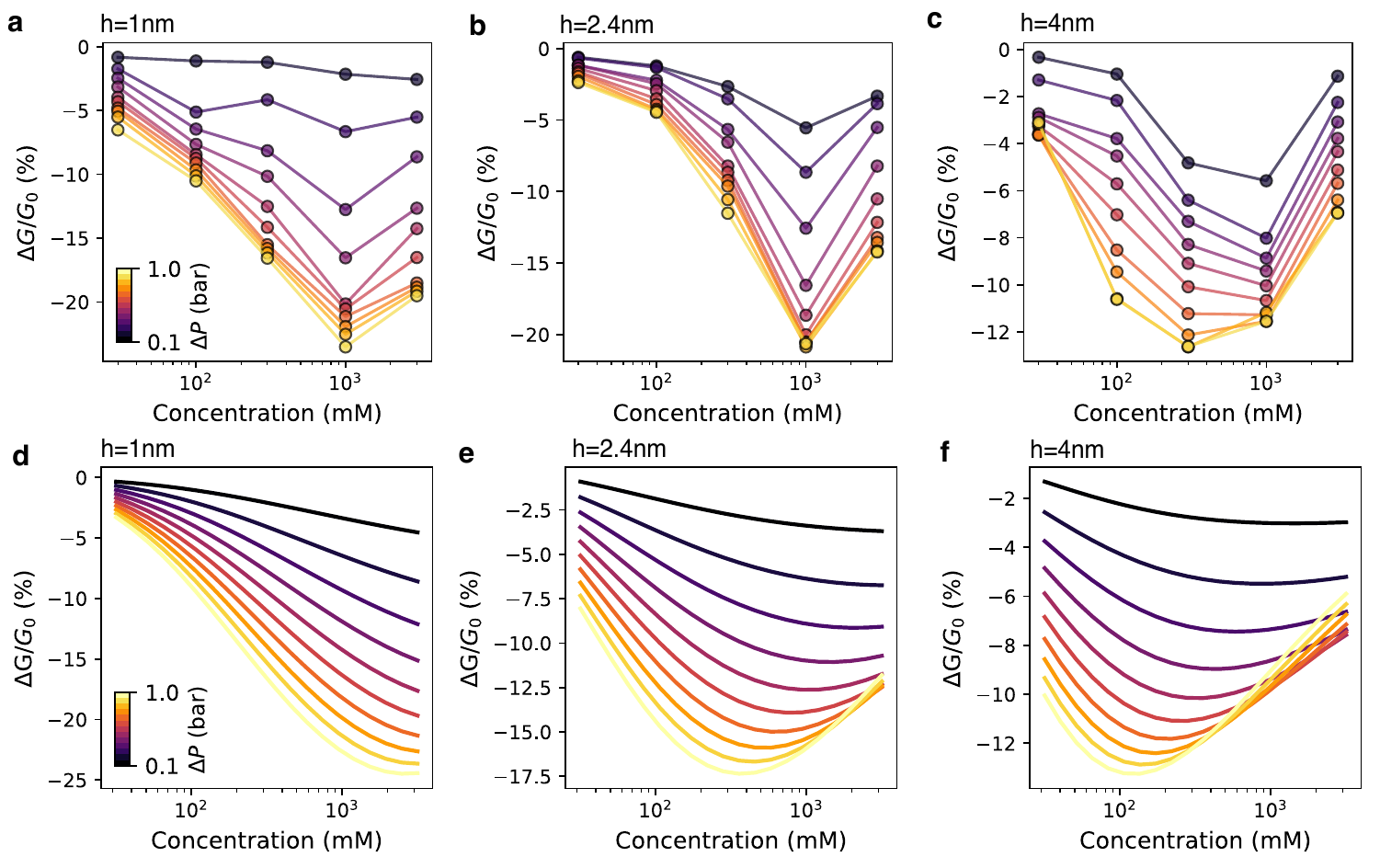}
	\caption{\textbf{{\color{black}Relative mechanosensitive response as a function of salt concentration for a range of pressure drops in channels of different heights.} }(\textbf{a-c}) Concentration-dependent mechanosensitivity curves for {\color{black} channels of different }height $h \in \{1, 2.4, 4\}$~nm for a set of pressures ranging between 0 and 1~bar. The same dataset is replotted in Supplementary Figure 9 as a function of pressure instead of concentration. \Rev{\textbf{(d-f)} Relative mechanosensitivity $\Delta G/G_0 (c_0,h,\Delta P)$ plotted according to the constrained one-dimensional EDL advection model without any free parameter. We take $b=20$~nm, $L=5$~$\mu$m, $\Sigma_{\rm inlet}=-200$~mC/m$^2 \times (c_0/1\textrm{M})^{0.7}$ and $\Sigma_{\rm ch}=-25$~mC/m$^2 \times (c_0/1\textrm{M})^{0.3}$}.}
	\label{figure2_revision}
\end{figure*}

\subsection*{Surface charge regulation}
We have shown above that the model predicts the suppression of the mechano-sensitive response for both vanishing and high salt concentrations. However, a quantitative investigation of the effect of salt concentration requires allowing for variations of the inlet surface charge with salinity $\Sigma_{\rm inlet}(c_0)$ due to a possible surface charge regulation \cite{secchi2016scaling}.

We report \Rev{in} Figure \ref{figure4}\textbf{a} the ${\Delta G/G_0}(\Delta P)$ curves of a channel of height $h=$ 2.4~nm \Rev{for a range of concentrations.} We fit the experimental data for all concentrations and confinements with the \Rev{the 1D EDL-advection model}. \Rev{Consistently with previous fits} at $c_0=1$~M, we fix the slip length to 20~nm, \Rev{assumed here to be independent of the salt concentration $c_0$ as a first approximation.} \LBB{Let us mention here that direct measurements of the slip length, with either nanoscale permeability \cite{secchi2016massive,qin2011measurement} or colloidal probe AFM \cite{lizee2024anomalous,maali2008measurement} measurements, are highly challenging and are still missing in 2D channels to our knowledge. New experiments are needed in particular to fully elucidate the dependence on confinement and salinity.} %\cite{lizee2024liquid}.} 
We further assume a concentration regulation of the channel's surface charge $\Sigma_{\rm ch} \sim c_0^{1/3}$ (with $\Sigma_{\rm ch} = -25$~mC/m$^2$ at 1~M) as a 1/3 exponent was previously reported for \Rev{the inner walls of carbon nanotubes which resembles pristine graphite} \cite{secchi2016scaling}. The only remaining free parameter is again the surface charge on the activated patch $\Sigma_{\rm inlet}$.

Overall, this single parameter model fits well the experimental curves across the entire range of concentrations and \Rev{channel-}heights (see Figure \ref{figure4}\textbf{a} and Supplementary Figure 9). \Rev{In} Figure \ref{figure4}\textbf{b}, we plot the fitted $\Sigma_{\rm inlet}$ as a function of salt concentration for all channels and find it to be \Rev{remarkably} independent of the channel's height. This highly constrained fit of the full dataset confirms the proper description of the mechanosensitivity measurements by \Rev{the 1D EDL-advection} model.

Although independent \Rev{of} $h$, $\Sigma_{\rm inlet}$ is found to depend on salt concentration \Rev{as a power law }$\Sigma_{\rm inlet}\propto c_0^\alpha$, and \Rev{$\alpha = 0.71 \pm 0.05$ here} (red dashed line). The regulation exponent we report here is higher than previously measured on carbon nanotubes \cite{secchi2016scaling} and other carbon surfaces \cite{emmerich2022enhanced}. Theoretical predictions of charge regulation do predict a scaling behavior of the surface charge $\Sigma_{\rm inlet}\propto c_0^\alpha$, with an exponent which can vary in the frame $\alpha \in [1/3;1]$ depending on conditions \cite{uematsu2018crossover}.

In Supplementary Note 2J, we propose a charge-regulation model extending on the description in Ref.\cite{secchi2016scaling} and accounting for a possible salt adsorption on the activated surface. This model predicts a $\Sigma_{\rm inlet}\propto c_0^{2/3}$ scaling,  close to our observations \Rev{$\alpha = 0.71 \pm 0.05$. This high value of $\alpha$ nevertheless still contrasts} with previous studies \cite{secchi2016scaling,emmerich2022enhanced}. We leave further investigations of such effects for future analysis. 

\subsection*{\Rev{Concentration-dependence in experiments \textit{versus} constrained EDL-advection model}}
\Rev{Now that all model parameters have been fully constrained, we plot the experimental measurements of $\Delta G/G_0$ against $c_0$ (in the range 30~mM to 3~M) for a range of pressure drops and channel heights $h\in\{1,2.4,4\}$~nm on Figure \ref{figure2_revision}\textbf{a-c} (data for the $h=7$~nm channel can be found in Supplementary Figure 9). For all channels, we find a non-monotonous trend with concentration : in absolute value, the mechanosensitivity vanishes at both low and high concentration and has a maximum which shifts downwards with increasing channel height.}
\LBB{In Figure \ref{figure2_revision}\textbf{d-f}, we then report the predicted results of the constrained model}
   %the experimental results, we plot the constrained model's solution 
   \Rev{-- without additional parameters --  (with $b= 20$~nm, $L=5$~$\mu$m, $\Sigma_{\rm ch} = -25$~mC/m$^2 \times (c_0/1~\textrm{M})^{0.3}$ and $\Sigma_{\rm inlet} = -200$~mC/m$^2 \times (c_0/1~\textrm{M})^{0.7}$). The model reproduces all the main features of the experimental data: the non-monotonous behavior with concentration, the downward-shift of the maximum with increasing slit height and the quantitative values for $\Delta G/G_0$ across parameters. Overall, the quantitative and qualitative agreement between the experimental results and the EDL-advection model confirms the accuracy of our modelling.} \LBB{Beyond, it illustrates how} \Rev{ engineering surface charge distributions allows designing an ionic, strain-free, nanofluidic pressure-sensor.}

In this work, we introduce a new class of mechanosensitive devices based on ion transport in asymmetrically charged nanochannels. Designing a surface charge pattern by exposing one graphite inlet to an oxygen plasma causes a modulation of ionic conductance under pressure drops. In contrast to most existing mechanisms, this effect informs on the sign of the pressure drop and does not rely on any structural deformation while maintaining mbar-range sensitivity, comparable to that of biological porins. Our theoretical model explains remarkably well the conductance-pressure response over a wide range of salt concentrations, including the strong enhancement in mechanosensitivity as the channel height drops to 1 nm. Our work opens up new avenues for the design of advanced nanofluidic sensing in solid state devices based on the precise spatial arrangement of surface charges for specific functions, as demonstrated in biological porins.

%In this study, we have demonstrated \LB{the emergence of} a new type of mechano-sensitive ion transport in two-dimensional graphite nanochannels, \LB{which can be tuned via the engineering of the surface pattern of the nanofluidic system. } We show that, in contrast to the majority of mechano-sensitive devices, this system does not rely on elastic deformations but rather on an asymmetric ion advection. This mechanism  It is also weakly dependent on salt concentration, allowing in-situ pressure sensing over the full range of concentrations with a mbar-range sensitivity. Furthemore, our theoretical model \LB{based on the intertwinned electrohydrodynamic ion and water transport does capture} the pressure-dependent mechano-conductance on the entire range of concentrations and channel heights. Our work opens up new avenues for the design of advanced nanofluidic sensing devices based on the precise spatial arrangement of surface charges for specific functions, as demonstrated in biological porins.

%\attention{Conclusion to be strengthened}

\RevEditor{\section*{Materials and Methods}}

\RevEditor{\subsection*{Device fabrication}}
\RevEditor{
The nanochannel fabrication procedure was refined based on previous studies, and a flow chart is presented in Supplementary Figure 1. First, we prepare the silicon nitride (SiN) membrane substrate. The commercially available SiN/Si wafer consists of two SiN layers on both sides of a 500~µm silicon (Si).  A square window was patterned on one side of the SiN layer by photolithography, followed by the removal of the SiN layer via reactive ion etching (RIE). The wafer was then immersed in a hot KOH solution ($70^\circ\text{C}$) for 12~h to etch through the middle Si and obtain a freestanding SiN membrane. After that, a rectangular slit was patterned and etched on the other side repeating the photolithography and RIE processes. Once the SiN substrate is ready, we then build the nanochannels on it. All graphite and graphene flakes used in this study are prepared using the mechanical exfoliation method on SiO$_2$ substrate.  First, a graphite crystal was transferred onto the SiN substrate using wet transfer method, serving as the bottom layer. After this, RIE was used from the backside of the SiN membrane to open the entrance to the channels. 
In parallel, we fabricated the spacer layer, which determines the dimensions of the channel. The spacer layer was patterned into tens of parallel strips using a combined electron beam lithography (EBL) and RIE,  with approximately 150~nm width (channel width) and separated by a similar distance. The cleaniness and height of the spacer layer were confirmed by atomic force microscopy (AFM), which corresponded to the height of the channel. Another graphite crystal was transferred onto the spacer via wet transfer method, serving as the top layer. 
This two-layer assembly consisting of the top and spacer layers was finally transferred onto the bottom crystal on the SiN substrate, ensuring that the channels were aligned perpendicular to the long edge of the rectangular slit. After each transfer, the sample undergoes a thorough cleaning with acetone and isopropanol, followed by annealing in an Ar/H2 atmosphere at 400°C for 3 hours to remove any potential contamination and polymer residue. Finally, a gold strip (3/50~nm Cr/Au) was deposited on top using lithography and e-beam evaporation to serve as a mask for RIE etching, which defines both the number and length of the channels. 
Prior to ionic transport measurements, the 'activated-inlet' nanochannel devices were treated with mild oxygen plasma (O$_2$/Ar with Moorfield Nanoetcher) for 1 minute. The O$_2$/Ar plasma treatment is performed using the nanoETCH Plasma Etching System from Moorfield Nanotechnology. The plasma treatment was carried out following a two-step recipe. (Step~1): RF power 12~W, duration 3~s, Ar 16~sccm, O$_2$ 8~sccm, pressure $2.6 \times 10^{-2}$~mbar. (Step~2): RF power 8~W, duration 60~s, with the same gas flows and pressure as in Step~1. Pristine-inlet nanochannel devices, on the other hand, are not plasma-treated. All fabrication processes were carried out in the clean room.}

\RevEditor{\subsection*{Ion transport measurements}}

\RevEditor{
For ion transport measurements, the channel device was used to separate two Teflon reservoirs filled with solutions of varying concentrations. The SiN/Si wafer with device under measurement was inserted between two half-cells and sealed with an O-ring to ensure that the nanochannels were the only paths between two half-cells. Two homemade Ag/AgCl electrodes were used as working electrodes to conduct ionic current. To prepare the Ag/AgCl electrode, a current of 1 mA was applied to a Ag wire (0.8 mm in diameter) for 30 min with voltage range of -1 to -2~V. The prepared Ag / AgCl electrode (dark brown) was then rinsed with deionized water and stored in the 1~M KCl solution. The potential difference between two Ag / AgCl electrodes in a 1~M KCl solution was measured to be less than 5~mV before they can be used for ionic conductance measurements in a range of electrolyte concentrations.
KCl solution of various concentration was used as the electrolyte. Measurements on the same device were conducted from low to high concentration. Between measurements, the cell was washed thoroughly with the solution of the new concentration before measuring. I–V curves were recorded using a Keithley 2636B SourceMeter controlled via custom in-house software. When all tests with different concentrations were completed, the cell and tubes were thoroughly washed with di-ionized water to remove any residues before moving to next samples. Pressure-driven streaming current measurements were performed using a precision pressure controller (FLUIGENT), connected to the fluidic cell and operated via OxyGEN software.
A constant bias voltage was applied across the nanochannel using a Keithley 2636B SourceMeter, while pressure steps were controlled through the software interface. Chronoamperometry (CA) was employed to record the ionic current (I(t)) response at each pressure step. The streaming current at each condition was extracted by averaging the steady-state current values, and the standard deviation was used to quantify measurement uncertainty.}

\RevEditor{\section*{Supplementary Materials}
Supplementary Materials for this article include Supplementary Notes 1, 2 and 3 providing respectively additional experimental results, a thorough derivation of the theoretical model and a detailed account of finite-element simulations.}

%\bibliography{pressure_bib}

\vspace{1em}
\section*{Acknowledgements} 
\RevEditor{
\textbf{Funding} The authors acknowledge support from ERC project n-AQUA, grant agreement $101071937$. B.C. acknowledges support from the NOMIS Foundation. ML acknowledges support from the Alexander von Humboldt foundation. \QY{ QY acknowledges support from the Royal Society University Research Fellowship URF$\backslash R1\backslash 221096$ and UK Research and Innovation Grant [EP/X017575/1]}. \textbf{Author contributions:} M.L, Q.Y and L.B conceived the project. Z.Z fabricated the nanochannel devices. Z.Z and M.L performed the experiments and analyzed the data. M.L, B.C and L.B developed the theoretical model. M.L performed the numerical calculations. M.L. and L.B. wrote the manuscript with inputs from all authors. All authors contributed to general discussion. \textbf{Competing interests:} The authors declare no competing interests.}

\subsection*{Data, Code, and Materials Availability}
\RevEditor{
All data needed to evaluate the conclusions in the paper are present in the paper and/or the Supplementary Materials. No new materials have been created for this work.}

\bibliography{pressure_bib}
\end{document}